\documentclass[twocolumn,showpacs,showkeys,preprintnumbers,amsmath,amssymb]{revtex4}
 \usepackage{dcolumn}
 \usepackage{bm}
\usepackage{graphicx}
 \begin{document}

 \title{Superradiant instability of the charged scalar field in
  stringy black hole mirror system}

 \author{Ran Li}

 \thanks{Electronic mail: liran.gm.1983@gmail.com}

 \author{Junkun Zhao}

 \affiliation{Department of Physics,
 Henan Normal University, Xinxiang 453007, China}

 \begin{abstract}

 It has been shown that the mass of the scalar field in the charged stringy black hole
 is never able to generate a potential well outside the event horizon
 to trap the superradiant modes. This is to say that
 the charged stringy black hole is stable against the massive charged
 scalar perturbation. In this paper we will study the superradiant instability
 of the massless scalar field in the background of charged stringy black hole
 due to a mirror-like boundary condition. The analytical expression of the
 unstable superradiant modes is derived by using the asymptotic matching method.
 It is also pointed out that the black hole mirror system becomes extremely unstable
 for a large charge $q$ of scalar field and the small mirror radius $r_m$.

 \end{abstract}



 \maketitle

 Long ago, there is proposal of building the black hole bomb
 \cite{press} by using the classical superradiance phenomenon
 \cite{zeldovich,bardeen,misner,starobinsky}.
 It seems that the mechanism of black hole bomb is very simple.
 When an impinging bosonic wave with the
 frequency satisfying the superradiant condition is scattered by the
 event horizon of the rotating black hole,
 the amplitude of this bosonic wave will be enlarged.
 If one places a mirror outside of the hole, the
 enlarged wave will be reflected into the hole once again.
 Then this wave will be bounced back and forth between the event horizon and the mirror.
 Meanwhile, the energy of this wave can become sufficiently big in this black hole mirror system until the mirror is destroyed.

 The black hole bomb mechanism firstly proposed by Press and Teukolsky \cite{press}
 was studied by Cardoso et. al. in \cite{cardoso2004bomb} recently.
 It is found that there exists a minimum
 mirror's radius to make the black hole mirror system unstable.
 See also the Refs.\cite{Rosa,Lee,leejhep,jgrosa,hod2013prd,hodbhb}
 for the recent studies on this topic.
 The black hole bomb mechanism can be generalized to other cases.
 The first case is to study the massive
 bosonic field in rotating black holes, for example in \cite{kerrunstable,detweiler,strafuss,dolan,
 Hod,hodPLB2012,konoplyaPLB,DiasPRD2006,zhangw,dolanprd2013},
 where the mass term can play the role of the reflecting
 mirror. In this case, the wave will be trapped in the potential well
 outside of the hole and the amplitude will grow exponentially, which
 triggers the instability of the system. The second case is to
 study the bosonic field perturbation in black hole background with
 the Dirichlet boundary condition at asymptotic infinity.
 These background spacetimes include the black holes in AdS spacetime \cite{cardoso2004ads,cardoso2006prd,KKZ,aliev,uchikata,rlplb,zhang},
 the black holes in G\"{o}del universe \cite{knopolya,rlepjc}, and the black hole in linear dilaton
 background \cite{clement,randilaton}. In all these spacetimes, the Dirichlet boundary condition
 provides the reflecting mirror, which results in the instabilities of the systems.

 For a charged scalar wave in the background of the spherical symmetric charged black hole,
 if the frequency of this impinging wave satisfying the superradiant condition,
 the wave will also undergo the superradiant process when scattered by the horizon \cite{bekenstein}.
 However, it is proved by Hod in \cite{hodrnplb2012,hodrnplb2013} that,
 for the Reissner-Nordstr\"{o}m (RN) black holes,
 the existence of a trapping potential well outside the black hole
 and superradiant amplification of the trapped modes cannot be satisfied simultaneously.
 This means that the RN black holes are stable under the perturbations of massive charged scalar fields.
 Soon after, Degollado et. al. \cite{Degolladoprd,Degollado} found that the same system can be made unstable by adding a
 mirror-like boundary condition like the case of the Kerr black hole.
 However, whether all of the charged black holes have the similar properties as the RN black hole
 is still an interesting question deserves further studies.

 In \cite{liprd}, we shown that the mass term of the scalar field in the charged stringy black hole
 is never able to generate a potential well outside the event horizon
 to trap the superradiant modes. This is to say that
 the charged stringy black hole is stable against the massive charged
 scalar perturbation. In this paper, we will further study the superradiant instability
 of the massless scalar field in the background of the charged stringy black hole
 due to a mirror-like boundary condition.

 This black hole is a the static spherical symmetric charged black holes
 in low energy effective theory
 of heterotic string theory in four dimensions, which is firstly found by
 Gibbons and Maeda in \cite{GM} and independently
 found by Garfinkle, Horowitz, and Strominger in \cite{GHS} a few years later.
 The metric is given by
 \begin{eqnarray}
 ds^2&=&-\left(1-\frac{2M}{r}\right)dt^2+\left(1-\frac{2M}{r}\right)^{-1}
 dr^2\nonumber\\
 &&+r\left(r-\frac{Q^2}{M}\right)(d\theta^2+\sin^2\theta d\phi^2)\;,
 \end{eqnarray}
 and the electric field and the dilaton field
 \begin{eqnarray}
 &&A_t=-\frac{Q}{r}\;,\nonumber\\
 &&e^{2\Phi}=1-\frac{Q^2}{Mr}\;.
 \end{eqnarray}

 The parameters $M$ and $Q$ are the mass and electric charge
 of the black hole respectively.
 The event horizon of black hole is located at $r=2M$.
 The area of the sphere of the charged
 stringy black hole approaches to zero when $r=Q^2/M$. Therefore, the
 sphere surface of the radius $r=Q^2/M$ is singular. When $Q^2\leq 2M^2$,
 this singular surface is surrounded by the event horizon.
 We will consider the black hole with the parameters
 satisfying the condition $Q^2\leq 2M^2$ in this paper.
 When $Q^2=2M^2$, the singular surface coincides with the
 event horizon. This is the case of extremal black hole.

 We start with analysing the scalar field perturbation in the background
 of the charged stringy black hole.
 The dynamics of the charged massless scalar field perturbation is governed
 by the Klein-Gordon equation
 \begin{eqnarray}
 (\nabla_\nu-iqA_\nu)(\nabla^\nu-iqA^\nu)\Psi=0\;,
 \end{eqnarray}
 where $q$ denote the charge of the scalar field.
 By taking the ansatz of the scalar field
 $\Psi=e^{-i\omega t}R(r)Y_{lm}(\theta,\phi)$,
 where $\omega$ is the conserved energy of the mode,
 $l$ is the spherical harmonic index, and $m$ is the
 azimuthal harmonic index with $-l\leq k\leq l$,
 one can deduce the radial wave equation in the form of
 \begin{eqnarray}
 \Delta\frac{d}{dr}\left(\Delta \frac{dR}{dr}\right)+UR=0\;,
 \end{eqnarray}
 where we have introduced a new function $\Delta=\left(r-r_+\right)\left(r-r_-\right)$
 with $r_+=2M$ and $r_-=Q^2/M$,
 and the potential function is given by
 \begin{eqnarray}
 U=\left(r-\frac{Q^2}{M}\right)^2(\omega r-qQ)^2-\Delta l(l+1)\;.
 \end{eqnarray}

 The classical superradiance phenomenon for the scalar field perturbation
 is present in charged stringy black hole \cite{dilatonsr}.
 In particular, by studying the asymptotic
 solutions of the radial wave equation near the horizon and
 at spatial infinity with the appropriate boundary conditions,
 one can obtain the superradiant condition of the charged scalar field \cite{liprd}
 \begin{eqnarray}
 \omega<q\Phi_H\;,
 \end{eqnarray}
 with $\Phi_H=\frac{Q}{2M}$ being the electric potential at the horizon.

 It has been shown by analyzing the behavior of the effective
 potential that for both the nonextremal black holes and the extremal black holes
 there is no potential well which is separated from the horizon by a potential barrier.
 Thus, the superradiant modes of charged scalar field can not be trapped
 and lead to the instabilities of the black holes.
 This indicates that the extremal and the nonextremal charged black holes in string theory
 are stable against the charged scalar field perturbations \cite{liprd}.

 In this paper, we will make the black hole unstable
 by placing a reflecting mirror outside of the hole.
 More precisely, we will impose the mirror's boundary condition that the scalar field
 vanishes at the mirror's location $r_m$, i.e.
 \begin{eqnarray}
 \Psi(r=r_m)=0
 \end{eqnarray}
 The complex frequencies satisfying the purely ingoing boundary at the black hole
 horizon and the mirror's boundary condition are called boxed quasinormal (BQN) frequencies \cite{cardoso2004bomb}.
 The scalar modes in the superradiant regime
 will bounce back and forth between event horizon and mirror.
 Meanwhile, the energy extracted from black hole by means of
 superradiance process will grow exponentially.
 This will cause the instability of the black hole mirror system.
 In the following, we will present an analytical calculations of BQN frequencies in a certain
 limit and show the instability in the superradiant regime caused by the mirror's boundary condition.

 Now we will employ the matched asymptotic expansion method
 \cite{page, unruh} to compute the
 unstable modes of a charged scalar field in this black hole mirror system.
 We shall assume that the Compton wavelength of the scalar particles
 is muck larger than the typical size of the black hole, i.e.
 $1/\omega\gg M$. With this assumption, we can divide
 the space outside the event horizon into two regions, namely, a near-region,
 $r-r_+\ll 1/\omega$, and a far-region, $r-r_+\gg M$. The approximated solution
 can be obtained by matching the near-region solution and the far-region solution
 in the overlapping region $M\ll r-r_+\ll 1/\omega$. At last,
 we can impose the mirror's boundary condition to obtain the
 analytical expression of the unstable modes in this system.

 Firstly, let us focus on the near-region in the vicinity of the event horizon,
 $\omega(r-r_+)\ll 1$. The radial wave function can be reduced to the form
 \begin{eqnarray}
 \Delta\partial_r(\Delta\partial_rR(r))
 +\left[(r_+-r_-)^2\varpi^2
 -l(l+1)\Delta\right]R(r)=0\;,
 \end{eqnarray}
 with the parameter $\varpi$ given by
  \begin{eqnarray}
 \varpi=r_+(\omega-q\Phi_H)\;.
 \end{eqnarray}
 Introducing the new coordinate variable
 \begin{eqnarray}
 z=\frac{r-r_+}{r-r_-}\;,
 \end{eqnarray}
 the near-region radial wave equation can be rewritten in the form of
 \begin{eqnarray}
 z\partial_z(z\partial_z R(z))
 +\left[\varpi^2-l(l+1)\frac{z}{(1-z)^2}\right]R(z)=0\;.
 \end{eqnarray}
 Through defining
 \begin{eqnarray}
 R=z^{i\varpi}(1-z)^{l+1}F(z)\;,
 \end{eqnarray}
 the near-region radial wave equation becomes the standard hypergeometric
 equation
 \begin{eqnarray}
 z(1-z)\partial_z^2F(z)+[c-(1+a+b)]\partial_zF(z)-abF(z)=0\;,
 \end{eqnarray}
 with the parameters
 \begin{eqnarray}
 a&=&l+1+2i\varpi\;,\nonumber\\
 b&=&l+1\;,\nonumber\\
 c&=&1+2i\varpi\;.
 \end{eqnarray}

 In the neighborhood of $z=0$, the general solution of
 the radial wave equation is then given in terms of the hypergeometric function
 \cite{handbook}
 \begin{eqnarray}
  R&=&Az^{-i\varpi}(1-z)^{l+1}F(l+1,l+1-2i\varpi,1-2i\varpi,z)
  \nonumber\\
  &&+Bz^{i\varpi}(1-z)^{l+1}F(l+1,l+1+2i\varpi,1+2i\varpi,z)\;.
 \end{eqnarray}
 It is obvious that the first term represents the ingoing wave
 at the horizon, while the second term represents the outgoing
 wave at the horizon. Because we are considering the classical
 superradiance process, the ingoing boundary condition at the
 horizon should be employed. Then we have to set $B=0$. The physical solution of
 the radial wave equation corresponding to the ingoing wave
 at the horizon is then given by
 \begin{eqnarray}
 R=Az^{-i\varpi}(1-z)^{l+1}F(l+1,l+1-2i\varpi,1-2i\varpi,z)\;.
 \end{eqnarray}

 In the far-region, $r-r_+\gg M$, the effects induced by the black hole
 can be neglected. The metric is reduced to be the Minkowski metric in the spherical
 coordinates. Then the radial wave equation reduces to the wave equation of a scalar field
 in the flat background
 \begin{eqnarray}
 \partial_r^2(rR(r))+\left[\omega^2-\frac{l(l+1)}{r^2}\right](rR(r))=0\;.
 \end{eqnarray}
 This equation can be solved by the Bessel function, which is given by \cite{handbook}
 \begin{eqnarray}
 R=r^{-1/2}\left[\alpha J_{l+1/2}(\omega r)+\beta J_{-l-1/2}(\omega r)\right]\;.
 \end{eqnarray}

 In order to match the far-region solution with the near-region
 solution, we should study the large $r$ behavior of the near-region solution
 and the small $r$ behavior of the far-region solution.
 For the sake of this purpose, we can
 us the $z\rightarrow 1-z$ transformation law for the hypergeometric function \cite{handbook}
 \begin{eqnarray}
 F(a,b,c;z)&=&\frac{\Gamma(c)\Gamma(c-a-b)}{\Gamma(c-a)\Gamma(c-b)}
 F(a,b,a+b-c+1;1-z)\nonumber\\
 &&+(1-z)^{c-a-b}
 \frac{\Gamma(c)\Gamma(a+b-c)}{\Gamma(a)\Gamma(b)}\nonumber\\
 &&\times
 F(c-a,c-b,c-a-b+1;1-z)\;\;.
 \end{eqnarray}
 By employing this formula and using the properties of hypergeometric function
 $F(a,b,c,0)=1$, we can get the large $r$ behavior of the near-region solution as
 \begin{eqnarray}\label{nearsolutionlarge}
 R&\sim& A\Gamma(1-2i\varpi)\left[\frac{(r_+-r_-)^{-l}\Gamma(2l+1)}
 {\Gamma(l+1)\Gamma(l+1-2i\varpi)}r^{l}\right.
 \nonumber\\&&\left.
 +\frac{(r_+-r_-)^{l+1}\Gamma(-2l-1)}
 {\Gamma(-l)\Gamma(-l-2i\varpi)}r^{-l-1}\right]\;.
 \end{eqnarray}

 On the other hand, using the asymptotic form of the Bessel function \cite{handbook},
 $J_\nu(z)=(z/2)^\nu/\Gamma(\nu+1)\;(z\ll 1)$, one can the small $r$ behavior
 of the far-region solution as
 \begin{eqnarray}
 R\sim \alpha \frac{(\omega/2)^{l+1/2}}{\Gamma(l+3/2)}r^l
 +\beta\frac{(\omega/2)^{-l-1/2}}{\Gamma(-l+1/2)}r^{-l-1}\;.
 \end{eqnarray}

 By comparing the large $r$ behavior of the near-region solution with
 the small $r$ behavior of the far-region solution, one can conclude
 that there exists the overlapping region $M\ll r-r_+\ll 1/\omega$
 where the two solutions should match. This matching yields
 the relation
 \begin{eqnarray}
 \frac{\beta}{\alpha}&=&\frac{\Gamma(-l+1/2)}{\Gamma(l+3/2)}
 \frac{\Gamma(l+1)}{\Gamma(2l+1)}
 \frac{\Gamma(-2l-1)}{\Gamma(-l)}
 \frac{\Gamma(l+1-2i\varpi)}{\Gamma(-l-2i\varpi)}\nonumber\\
 &&
 \times\left(\frac{\omega}{2}\right)^{2l+1}(r_+-r_-)^{2l+1}\;.
 \end{eqnarray}
 By using the property of Gamma function, $\Gamma(x+1)=x\Gamma(x)$,
 one can easily derive these relations
 \begin{eqnarray}
 &&\frac{\Gamma(-l+1/2)}{\Gamma(l+3/2)}=\frac{(-1)^l 2^{2l+1}}{(2l-1)!!(2l+1)!!}\;,\nonumber\\
 &&\frac{\Gamma(-2l-1)}{\Gamma(-l)}=\frac{(-1)^{l+1}l!}{(2l+1)!}\;,\nonumber\\
 &&\frac{\Gamma(l+1-2i\varpi)}{\Gamma(-l-2i\varpi)}
 =(-1)^{l+1}2i\varpi \prod_{k=1}^{l}(k^2+4\varpi^2)\;.
 \end{eqnarray}
 Applying these formulas into the matching condition, one can derive
 \begin{eqnarray}
 \frac{\beta}{\alpha}&=&2i\varpi \frac{(-1)^l}{(2l+1)}
 \left(\frac{l!}{(2l-1)!!}\right)^2
 \frac{(r_+-r_-)^{2l+1}}{(2l)!(2l+1)!} \nonumber\\
 &&\times \prod_{k=1}^{l}(k^2+4\varpi^2) \omega^{2l+1}\;.
 \end{eqnarray}

 Now we want to impose the mirror's boundary condition to study the
 unstable modes. We assume that the mirror is placed near the infinity
 at a radius $r=r_m$. The far-region radial solution should vanish
 when reflected by the mirror. This yields the extra condition between
 the amplitudes $\alpha$ and $\beta$ of the far-region radial solution,
 which is given by
 \begin{eqnarray}
 \frac{\beta}{\alpha}=-\frac{J_{l+1/2}(\omega r_m)}{J_{-l-1/2}(\omega r_m)}\;.
 \end{eqnarray}

 This mirror condition together with the matching condition give us the following equation
 which determines the BQN frequencies of the scalar field in this black hole mirror system
 \begin{eqnarray}
 \frac{J_{l+1/2}(\omega r_m)}{J_{-l-1/2}(\omega r_m)}
 &=&2i\varpi \frac{(-1)^{l+1}}{(2l+1)}
 \left(\frac{l!}{(2l-1)!!}\right)^2
 \frac{(r_+-r_-)^{2l+1}}{(2l)!(2l+1)!}\nonumber\\
 &&\times \prod_{k=1}^{l}(k^2+4\varpi^2) \omega^{2l+1}\;.
 \end{eqnarray}

 For the very small $\omega$, the analytical solution of BQN frequencies
 can be found from the above relation. In this case, the right hand side of the
 above relation is very small and then can be set to be zero.
 This means that
 \begin{eqnarray}
 J_{l+1/2}(\omega r_m)=0\;.
 \end{eqnarray}
 The real zeros of the Bessel functions were well studied. We shall label the
 $n$th positive zero of the Bessel function $J_{l+1/2}$ as $j_{l+1/2,n}$. Then
 we can get
 \begin{eqnarray}
 \omega r_m=j_{l+1/2,n}\;.
 \end{eqnarray}
 In the first approximation for BQN frequencies, the solution of the eq.(26)
 has a small imaginary part, which can be written as
 \begin{eqnarray}
 \omega_{BQN}=\frac{j_{l+1/2,n}}{r_m}+i\delta\;,
 \end{eqnarray}
 where the introduced imaginary part $\delta$ is small enough comparing the
 real part of BQN frequency.It can be considered as a correction to eq.(28).
 For the small $\delta$, we can use the Taylor expansion
 of Bessel function $J_{l+1/2}(\omega r_m)=i\delta r_m J'_{l+1/2}(j_{l+1/2,n})$.
 Then the equation (26) can be reduced to
 \begin{eqnarray}
 i\delta r_m
 \frac{J'_{l+1/2}(j_{l+1/2,n})}{J_{-l-1/2}(j_{l+1/2,n})}
 =2i\varpi \frac{(-1)^{l+1}}{(2l+1)}
 \left(\frac{l!}{(2l-1)!!}\right)^2&&\nonumber\\
 \times\frac{(r_+-r_-)^{2l+1}}{(2l)!(2l+1)!}
  \prod_{k=1}^{l}(k^2+4\varpi^2) \left(\frac{j_{l+1/2,n}}{r_m}\right)^{2l+1}\;.&&
 \end{eqnarray}
 From this we can easily obtain the small imaginary part of the BQN frequencies as
 \begin{eqnarray}
 \delta=-\gamma \left(\frac{j_{l+1/2,n}}{r_m}-q\Phi_H\right) \frac{(-1)^lJ_{-l-1/2}(j_{l+1/2,n})}{J'_{l+1/2}(j_{l+1/2,n})}\;,
 \end{eqnarray}
 with
 \begin{eqnarray}
 \gamma=\frac{2}{(2l+1)}\left(\frac{l!}{(2l-1)!!}\right)^2
 \frac{r_+(r_+-r_-)^{2l+1}}{r_m(2l)!(2l+1)!}&&\nonumber\\
  \times\left(\prod_{k=1}^{l}(k^2+4\varpi^2)\right)
  \left(\frac{j_{l+1/2,n}}{r_m}\right)^{2l+1}\;.&&
 \end{eqnarray}
 Notice that $\gamma$ is always greater than zero,
 and $(-1)^lJ_{-l-1/2}(j_{l+1/2,n})$ and $J'_{l+1/2}(j_{l+1/2,n})$
 always have the same sign. So we have
 \begin{eqnarray}
 \delta \propto -(\textrm{Re}[\omega_{BQN}]-q\Phi_H)\;.
 \end{eqnarray}
 It is easy to see that, in the superradiance regime,
 $\textrm{Re}[\omega_{BQN}]-q\Phi_H<0$,
 the imaginary part of the complex BQN frequency
 $\delta>0$. The scalar field has the time dependence
 $e^{-i\omega t}=e^{-i \textrm{Re}[\omega] t}e^{\delta t}$, which implies
 the exponential amplification of superradiance modes.
 This indicates that the BQN frequencies in the superradiant regime
 is unstable for the charged scalar field in the stringy black hole with a
 mirror placed outside of the hole.

 Here, we shall discuss our analytical result briefly.
 The instability time scaling that characterizes the composed
 black hole mirror system is given by
 \begin{eqnarray}
 \tau=\frac{1}{\delta}\;.
 \end{eqnarray}
 Firstly, the imaginary part of the complex BQN frequency
 $\delta$ decreases when the mirror's radius $r_m$ increases.
 This means that the instability time scaling becomes larger
 for the larger mirror radius.

 Secondly, from equation (29), we can observe that
 wave frequencies of these unstable superradiant modes
 is propotional to the inverse of the mirror radius. When the mirror
 radius decreases, the allowed wave frequencies will increase. The superradiant
 condition then restricts that the mirror can not be placed very near the
 horizon. There exist a critical radius $r_m^{crit}$ at which this instability disappears.
 From the analytical result, one can obtain the critical radius is given by
 \begin{eqnarray}
 r_m^{crit}=\frac{j_{l+1/2,n}}{q\Phi_H}\;.
 \end{eqnarray}
 However, from the above equation, one can see that we can still place the mirror
 at a very small radius as long as the charge $q$ of the scalar field
 is big enough.

 At last, one can also observe that $\delta$ grows with
 the charge $q$ of the scalar field. This implies the instability becomes more
 stronger as $q$ increases. So one can expect that, for the large $q$ and small
 $r_m$, the instability time scale of this charge spherical symmetric black hole mirror system
 will become very short. This result is different from the rotating
 black hole mirror system. For the rotating black hole \cite{cardoso2004bomb},
 the superradiant condition is given by $\omega<m\Phi_H$, where $m$ and $\Phi_H$
 are the azimuthal number and the angular velocity of the horizon, respectively.
 The value of $m$ can not be taken arbitrarily large because of the limit condition
 $m\leq l$ with $l$ being the spherical harmonic index.

 In summary, we have studied the instability
 of the massless charged scalar field in the stringy black hole mirror system.
 By imposing the mirror boundary condition, we have analytically calculated
 the expression of BQN frequencies. Based on this result,
 we also point out that the black hole mirror system becomes extremely unstable
 for the large charge $q$ of scalar field and the small mirror radius $r_m$.
 In \cite{hod2013prd}, it is deduced by Hod using the analytical method that,
 for the RN black hole, the instability time scale can be made arbitrary short in special limit.
 So, the analytical computation and the numerical simulation are
 still required to verify the conclusion.

 \section*{ACKNOWLEDGEMENT}
 This work was supported by NSFC, China (Grant No. 11205048).

 \end{document}